\documentclass[11pt]{article}
\usepackage{amsmath}
\usepackage[english]{babel}
\usepackage{setspace,amsbsy,amssymb,physics,cite,titlesec,graphics}
\usepackage{hyperref}
\usepackage[nameinlink]{cleveref}
\usepackage[left=2.5cm,right=2.5cm,top=2.5cm,bottom=2.5cm]{geometry}
\def\LL{\mathcal L}
\def\PP{\mathcal P}
\def\CC{{\mathcal C}}
\def\RR{{\mathcal R}}
\def\KK{{\mathcal K}}
\def\MM{{\mathcal M}}
\def\ZZ{{\mathcal Z}}
\def\OO{\mathcal{O}}
\def\XX{\mathcal{X}}
\def\VV{\mathcal V}
\def\SS{\mathcal S}
\def\io{{\mathrm i}}
\newcommand{\be}{\begin{equation}}
\newcommand{\ee}{\end{equation}}
\newcommand{\bes}{\begin{split}}
\newcommand{\ees}{\end{split}}
\newcommand{\sbkt}[1]{\left[#1\right]}
\newcommand{\bkt}[1]{\left(#1\right)}
\newcommand{\p}{\partial}
\newcommand{\SO}[1]{\mathrm{SO}(#1)}
\newcommand{\U}[1]{\mathrm{U}(#1)}
 \hypersetup{colorlinks=true}
\hypersetup{linkcolor=black}
\hypersetup{citecolor=red}
\hypersetup{urlcolor=blue}
\makeatletter
\renewcommand{\@dotsep}{1000} 
\makeatother
 \numberwithin{equation}{section}
\titleformat{\section}{\normalfont\bfseries}{\thesection.}{4pt}{}
\titlespacing{\section}{0pt}{25pt}{6pt}
\titleformat{\subsection}{\normalfont\itshape}{\thesubsection.}{4pt}{}
\titlespacing{\subsection}{0pt}{15pt}{6pt}
\titleformat{\subsubsection}{\normalfont\itshape}{\thesubsubsection.}{4pt}{}
\titlespacing{\subsubsection}{0pt}{15pt}{6pt}
\DeclareFontShape{OT1}{cmr}{mx}{n}
    {<->cmr10}{}
\newcommand{\mytitlefont}{\fontseries{mx}\selectfont}
\DeclareMathAlphabet{\titlemath}{OT1}{cmr}{mx}{n}
\begin{document}
\onehalfspacing
%
% Titlepage
%
\begin{titlepage}
\thispagestyle{empty}
\begin{flushright} \small
 UUITP-04/20\\
 \end{flushright}
\smallskip

\begin{center}

~\\[2cm]

{\fontsize{26pt}{0pt} \mytitlefont 
Entanglement Entropy \\ in Closed String Theory
 }

~\\[0.5cm]
 Usman Naseer
~\\[0.1cm]
 {\it 
Department of Physics and Astronomy, Uppsala University, 
\\
Box 516, SE-751 20 Uppsala, Sweden
 }
 ~\\[0.15 cm]
 {\texttt
 usman.naseer@physics.uu.se
 }
~\\[0.8cm]

\end{center}
%
%Abstract
%  
\abstract{\noindent
In local quantum field theory, the entanglement entropy of a region is divergent due to the arbitrary short-wavelength correlations near the boundary of the region. Quantum gravitational fluctuations are expected to cut off the entropy of the ultraviolet modes. Guided by this, we study the entanglement entropy in closed string theory. We factorize the configuration space of closed strings based on the position of the center of mass. We then compute the one-loop R\'enyi partition functions corresponding to this factorization using the replica method and string field theory. The short-wavelength modes are cut off at the string scale and the one-loop entanglement entropy is ultraviolet-finite.  A non-trivial path integration measure, required to produce the correct one-loop vacuum amplitude, plays a key role.
 }
\vfill
%
%Date
%
\begin{flushleft}
\today
\end{flushleft}
\end{titlepage}
%
%Table of contents
%
\tableofcontents
\baselineskip15pt
%
%body of the article
%
%
\section{Introduction}
Over the last decade concepts from quantum information theory have given remarkable insights into the structure of quantum 
field theory and quantum gravity. An important quantity in this context is the entanglement entropy 
(see~\cite{Nishioka:2018khk,Headrick:2019eth}
 and references therein for a review). In a local quantum field theory with a background spacetime, the entanglement entropy has UV divergences due to the correlations between degrees of freedom near the entangling surface. This leads to the 
so-called `area-law' of entanglement entropy for typical vacuum states
\be\label{eq:alaw}
S= c_0 {A\over \epsilon^{D-2}}+\cdots+S_{\rm finite}.
\ee
The leading divergence in entanglement entropy is proportional to the area $A$ of the entangling surface. Here $D$ is the 
spacetime dimension, $\epsilon$ is the UV-cutoff and $c_0$ is a constant which depends on the details of the regularization 
scheme. `$\cdots$' denotes subleading divergences. The finite part $S_{\rm finite}$ depends on the regularization scheme. In even dimensions there are also logarithmic divergences whose coefficients are regularization scheme independent.

 Arguments based on black hole physics, however, suggest that the entanglement entropy is finite in a quantum theory of 
gravity~\cite{Susskind:1994sm,Larsen:1995ax,Jacobson:1994iw}. The basic idea is the following (see~\cite{Wall:2018ydq} for a recent review of 
related issues). In a black hole background with matter outside the horizon, the total entropy must include contributions from the 
black hole entropy $S_{\rm BH}={A\over 4 G}$ and the entropy $S_{\rm out}$ of the fields outside the horizon~\cite{Hawking:1974sw,Bekenstein:1972tm}.  $S_{\rm BH}$ 
depends on the UV-cutoff via the renormalization of the gravitational coupling constant while $S_{\rm out}$ depends on the UV-cutoff as in~\cref{eq:alaw}. If the same regularization scheme is 
used to compute $S_{\rm out}$ and the renormalized Newton's constant then  divergences in $S_{\rm out}$ match the ones that renormalize Newton's constant. More concretely 
\be
{A\over 4 G_{\rm bare}} +c_0{ A\over \epsilon^{D-2}}= {A\over 4 G_{\rm ren}}.
\ee
So the total entropy $S_{\rm BH}+ S_{\rm out}$ is renormalization group invariant.
Gravity 
becomes strongly coupled at the Planck scale (or the smallest possible length scale in quantum gravity) and ${1\over G}\to 0$. 
At the Planck scale total entropy is just the entanglement entropy which has become finite due to quantum gravitational effects. The partitioning of the total entropy into $S_{\rm BH}$ and $S_{\rm out}$ and their behavior under the renormalization 
group was studied in detail in~\cite{Jacobson:2012ek,Cooperman:2013iqr} providing further evidence that the entanglement 
entropy is a well-defined observable in quantum gravity. Indeed assuming this finiteness was crucial to derive the Einstein's 
equations from the maximal vacuum entanglement hypothesis~\cite{Jacobson:2015hqa}. Motivated by these ideas we study  
entanglement entropy in the closed  string theory.
   
A generic tool to compute entanglement entropy in field theory is the replica method. This requires computing the partition 
function of the theory on an $n$-fold branched cover of the spacetime and then analytically continuing in $n$. An important obstacle 
in the application of this method to the worldsheet string theory is that there is no known conformal field theory with a target 
space which is a branched cover. Nevertheless, in  earlier works~\cite{Susskind:1994sm,Dabholkar:1994ai,He:2014gva}, it was 
proposed that one can use a $\mathbb{Z}_N$ orbifold with the formal identification $n={1\over N}$ as a background and then 
analytically continue in $N$. Recently,  
 subtleties regarding this approach for open strings are discussed and clarified in~\cite{Witten:2018xfj}.  Such progress had 
remained elusive for the case of closed string theory: it is not known how to write the orbifold partition function as an analytic 
function of $N$.\footnote{This should be contrasted with the thermodynamic entropy which can be computed by studying string 
theory on a background where the time direction is a circle with radius proportional to the inverse 
temperature~\cite{Dabholkar:1994gg,Polchinski:1985zf,Obrien19871184,McClain1987}. 
Since the sigma model with a compact time direction in the target space is 
a conformal field theory one does not encounter the subtleties associated with the computation of the entanglement entropy.}   
 
 An alternative approach is to study entanglement properties using the formalism of string field theory (SFT). The framework of 
SFT was developed to obtain non-perturbative insights into the dynamics of strings~\cite{Schnabl:2005gv}. In recent years it has 
become clear that a consistent perturbative formulation of string theory also needs the framework of SFT. Two instructive 
examples are mass renormalization~\cite{Pius:2013sca,Pius:2014iaa} and the vacuum shift~\cite{Pius:2014gza} which  cannot 
be addressed  using conventional methods based on a worldsheet conformal field theory. Worldsheet conformal invariance 
imposes the tree level on-shell conditions and vacuum expectation values. Loop-corrections can change the physical mass and 
generate new terms in the potential so that the new vacuum does not satisfy the classical equations of motion and hence cannot 
be described by a worldsheet conformal field theory. Nevertheless, mass renormalization and vacuum shift are standard 
problems in quantum field theory and SFT provides a systematic procedure to address these. The ability of SFT to deal with 
off-shell 
backgrounds makes it a promising framework to study the entanglement entropy in string theory. In the language 
of~\cite{Susskind:1994sm}, SFT provides a prescription to compute the off-shell generating functional which is needed for replica 
method. Since entanglement entropy is a physical quantity one expects it to be independent of the prescription.  
  
Indeed, the entanglement entropy for open strings was studied using SFT in~\cite{Balasubramanian:2018axm}. The result was found to be 
consistent with the effective field theory: the entanglement entropy is equal to the sum of entanglement entropies of all fields in 
the spectrum of open string theory. This is compelling evidence that the SFT provides a suitable formalism to study entanglement entropy. 
 In this paper we use free light-cone SFT to study the entanglement entropy in closed string theory.

In a theory where fundamental degrees of freedom are non-local objects like strings, an important issue is to appropriately define entanglement entropy. In an ordinary quantum field theory the definition of entanglement entropy is based on the inherent locality of the theory. Given a subspace of the Cauchy surface, the degrees of freedom naturally factorize into those that are inside the subspace and the ones that are outside. After tracing out the degrees of freedom outside, one obtains reduced density matrix for the degrees of freedom inside. Entanglement entropy is then defined to be the von Neumann entropy of the reduced density matrix.  Since string theory describes extended objects we do not have simple notion of locality.
Apart from the strings completely inside or outside a given subspace, one also has strings which are partially inside and partially outside the subspace. This leads to various subtleties regarding the definition of subregions and entanglement entropy. To factorize the degrees of freedom into two parts, one needs additional information. Note that this subtlety is present for open and closed strings alike. One simple choice is to factorize the degrees of freedom based on the position of the center-of-mass of the string as proposed in~\cite{Balasubramanian:2018axm}. We divide the space of strings in two parts: the ones with center-of-mass inside the subspace and the ones with center-of-mass outside the subspace. Given this factorization, one can proceed as in the case of the \emph{ordinary} field theory and compute entanglement entropy. 

We emphasize that there are other choices of factorization which will give different definitions of entanglement entropy. But it is the definition proposed by~\cite{Balasubramanian:2018axm} that we use to compute entanglement entropy in closed string theory and show that it is UV-finite. More refined definitions may lead to different quantitative results but we expect the qualitative results to be the same. 

 At a technical level, the mechanism 
responsible for the finiteness is the same which makes the one-loop vacuum amplitude finite in closed string theory. The 
one-loop 
vacuum amplitude involves an integration over the moduli space of the torus. It avoids the problematic regions which give 
rise to UV-divergences in a quantum field theory. The UV-finite vacuum amplitude is obtained in SFT by a careful analysis of the path integral measure and the field basis~\cite{Sen:1993kb}. 
The same path integral measure and the field basis then also give UV-finite  R\'enyi partition functions and the entanglement entropy\footnote{
Since the one-loop 
vacuum amplitude is independent of the string coupling our result should be understood as computing the one-loop contribution 
to vacuum entanglement entropy. At finite coupling one would also expect a `classical' contribution which is proportional to 
$\frac{1}{g_s^2}$ 
and is expected to produce the  ${\text{Area}\over 4 G_N}$ term. We are not able to compute this contribution using our methods. Perhaps covariant SFT and the methods of~\cite{Zwiebach:1996ph} can be used to address this important question.  
We thank Raghu Mahajan and a referee for discussions on this point.}.

This paper is organized as follows: In~\cref{sec:setup} we discuss and review aspects of the replica method to compute entanglement entropy in generic field theories. In~\cref{sec:EEinOSFT} we use the replica method to reproduce the results of~\cite{Balasubramanian:2018axm} for open SFT. In~\cref{sec:EEinCSFT} we introduce basic elements of closed SFT and compute entanglement entropy. In~\cref{sec:conclusion} we conclude with a discussion of various subtleties in our definition of entanglement entropy and furture directions. 
\section{Preliminaries}\label{sec:setup}
\subsection{Replica method}\label{subsec:replica}
In this section we review some basic facts about the entanglement entropy and the replica method in quantum field theories. We 
start by
considering a scalar field theory on `spacetime' $\MM$ of dimension $d_\MM$ described by the action 
\be
{\rm I}=\int d^{d_\MM} x\,  \phi\  \mathcal{O}\bkt{\partial_\mu,x^\mu}\phi.
\ee
We choose $x^\mu, \mu=0,1,\cdots, d_\MM-1$ as coordinates on $\MM$ and $x^0$ is treated as Euclidean time direction. 
 $\mathcal{O}\bkt{\partial_\mu,x^\mu}$ is a positive semi-definite operator whose eigen-functions span the function space on 
$\MM$. 
In general $\OO\bkt{\p_\mu,x^\mu}$ can involve derivatives, explicit position dependence and other constant parameters 
such as length or mass scales. The zero-modes of $\OO\bkt{\p_\mu,x^\mu}$ are on-shell field configurations for which the action 
is zero. The partition function of the theory can be computed by expanding the field in eigen-functions of 
$\OO\bkt{\p_\mu,x^\mu}$ 
and then performing the Gaussian integration over non-zero modes. This results into 
\be
\ZZ= \bkt{\det \OO\bkt{\p_\mu,x^\mu}}^{-\frac12},
\ee
where the determinant only involves positive eigenvalues of $\OO\bkt{\p_\mu,x^\mu}$. This can also be written as
\be
\log \ZZ=  \int_0^\infty\frac{dt}{2 t}\Tr  \KK_{\OO} \bkt{t},
\ee
where $\KK_\OO\bkt{t,x:x'}$ is the heat kernel associated with the operator $\OO\bkt{\p_\mu,x^\mu}$, i.e., 
\be
\langle  x|e^{-t\OO}|x'\rangle \equiv \KK_\OO\bkt{t,x:x'},\qquad \Tr\KK_\OO\bkt{t}=\int d^{d_\MM} x\  \KK_\OO\bkt{t,x:x}. 
\ee

To study the spatial entanglement of a given state in this theory we proceed as follows:  We choose a codimension-$1$ `Cauchy' 
surface $\Sigma$ inside $\mathcal{M}$  and partition it into two regions $\RR$ and its complement $\RR^c$:  $\Sigma=\RR \cup \RR^c$,  where $\RR$ is the region of 
interest. Given a state described by a density matrix, we then trace out the degrees of freedom in $\RR^c$ to get the reduced 
density matrix\footnote{Strictly speaking, this is not well-defined in a local quantum field theory because of the type-III 
property of the algebra of observables~\cite{Witten:2018lha}. We do not worry about such subtleties here as in practice this 
procedure leads to physically interesting results.} on $\RR$.  Entanglement entropy is then just the von Neumann entropy of the reduced 
density matrix. A convenient way to carry out this procedure for the vacuum state is the replica method.  We place the theory on 
$\mathcal{M}_n$  which is an $n$-fold branched cover of $\MM$. The branching is along the codimension-2 entangling surface  
$\p \RR$. The $n$'th R\'enyi partition function $\ZZ\bkt{n}$ is the partition function of the theory on $\MM_n$  and is proportional to $\Tr \rho^n_0$, where $\rho_0$ is the 
reduced density matrix of the vacuum state. Entanglement entropy is then computed by analytically continuing in $n$
\be
S= -\lim_{n\to 1}\bkt{n \p_n-1}\log \ZZ\bkt{n}=-\lim_{n\to 1} \bkt{n\p_n-1} \int_0^\infty {dt\over 2t}\Tr\KK_{\OO}^{(n)} \bkt{t} ,
\ee
where $\KK_\OO^{\bkt{n}}\bkt{t,x:y}$ is the appropriate heat-kernel on $\MM_n$.

In this paper, we restrict to Cauchy surfaces given by a constant time slice and the region $\RR$ given by the half space, i.e., 
\be\label{eq:SnR}
\Sigma=\{{x\in \MM} | x^0=0\},
\qquad
\RR=\{ x \in \Sigma | x^1 \geq 0\}.
\ee
The relevant branched cover $\MM_n$ has a conical singularity in the $\bkt{x^0, x^1}$-plane. If we parameterize this plane using 
polar coordinates $\bkt{r,\phi}$ then the polar angle is $2\pi n$ periodic on $\MM_n$. If the theory has $\SO{2}\sim\U{1}$ 
symmetry in the $(x^0,x^1)$-plane\footnote{This means that $\MM$ has an isometry which rotates $\bkt{x^0,x^1}$-plane and the 
kinetic operator $\OO\bkt{\p_\mu,x^\mu} $ is also invariant under the rotation.}  then  the heat kernel $\KK_\OO\bkt{t, x:x'}$  depends on 
the difference of the polar angles $\phi-\phi'$ and is $2\pi$ periodic. The heat kernel on $\MM_n$ is $2\pi n$ periodic and is given 
by~\cite{Fursaev:2011zz}
\be\label{eq:hkonMq}
\KK_\OO^{\bkt{n}}\bkt{t, \phi-\phi'}
=\KK_{\OO}\bkt{t,\phi-\phi'} + {1\over 4\io \pi n} \int_\CC
dz \cot\bkt{z\over 2 n} \KK_\OO\bkt{t, \phi-\phi'+z},
\ee
 where the contour $\CC$ consists of two vertical lines: the first one going from $-\pi-\io \infty$ to $-\pi+\io \infty$  and the second 
going from $\pi+\io\infty$ to $\pi-\io\infty$. In the above expression, we have suppressed the dependence of the heat kernel on 
all coordinates except the polar angle.
The entanglement entropy can then be computed by finding the trace of the heat kernel and analytically continuing in $n$. For a 
Lorentz invariant theory, i.e., $\MM=\mathbb{R}^D$ and $\OO\bkt{\p_\mu,x^\mu}=\OO\bkt{\p_\mu\p^\mu}$ we give a detailed 
derivation in \cref{app:ancont} where we show that the partition function on $\MM_n$ is given by
\be
\log \ZZ\bkt{n}= {
  A\over 12\bkt{4\pi}^{\tfrac D2-1}
 \Gamma\bkt{\tfrac D2-1}}
 {1-n^2 \over n}
\int_0^\infty {dt\over t}\int_0^\infty dp\, p^{D-3} e^{-t \OO\bkt{-p^2}}+\cdots.
\ee
The `$\cdots$' represents a term proportional to $n$ which does not contribute to entanglement entropy 
 \be\label{eq:EEhalfLI}
S
={ A \over  6 \bkt{4\pi}^{\tfrac D2-1} \Gamma\bkt{\tfrac D2-1}}
\int_0^\infty {dt\over t}
\int_0^\infty dp\, p^{D-3} e^{-t \OO\bkt{-p^2}}.
 \ee
Here $A=\int d^{D-2} x$ is the area of the entangling surface $\p \RR$. For generic Lorentz-invariant kinetic terms, the integral 
over $t$ diverges which leads to the divergent entanglement entropy:   $S\sim {A\over \epsilon^{D-2}}$, where $\epsilon$ is a 
UV-cutoff.

Now consider the theory on a spacetime of the form $\MM=\mathbb{R}^D\times \XX$,  where $\XX$ is some 
$d_\XX$-dimensional 
space. The fields now depend on the coordinates $x^\mu$ on $\mathbb{R}^D$ and coordinates along $\XX$ which 
we do not specify. We take the kinetic operator to be $\OO\bkt{\p_\mu\p^\mu}+\OO_\XX$ where $\OO_\XX$ is some differential operator on $\XX$. The heat kernel of the sum of the operator is a product of the heat kernels for the two operators. 
 If we are interested in computing the entanglement entropy of  the region defined in~\cref{eq:SnR}, i.e., Cauchy surface and the subregion are defined by imposing appropriate conditions on $\mathbb{R}^D$, then the heat kernel on the 
relevant branched cover also factorizes.
 \be
 \KK_{\OO+\OO_\XX}^{\bkt{n}}= \KK_\OO^{\bkt{n}}\ \KK_{\OO_\XX}.
 \ee
 Upon analytically continuing in $n$ we see that the entanglement entropy of the half space for the theory on 
$\MM=\mathbb{R}^D\times 
\XX$ is 
 \be
S
=\frac{ A}{ 6 \bkt{4\pi}^{\tfrac D2-1}} 
\frac{1}{\Gamma\bkt{\frac{D}{2}-1}}
\int_0^\infty {dt\over t}
\int_0^\infty dp\, p^{D-3} e^{-t \OO\bkt{-p^2}}\Tr  \KK_{\OO_\XX}\bkt{t}.
 \ee
 In the case where the operator $\OO\bkt{-p^2}={p^2\over 2}$, i.e., the usual kinetic operator, the above expression simplifies to 
 \be\label{eq:EEhalfLI}
 S
= {A \over 12 \bkt{2\pi}^{\tfrac D2-1}}
\int_0^\infty {dt\over t^{\tfrac D2}}
\Tr  \KK_{\OO_\XX}\bkt{t}.
 \ee
 
 Let us now comment on how the methods reviewed above apply to string theory. SFT is defined on the space of configurations of strings. Configuration of a string is specified by its center-of-mass and an infinite number of oscillation modes. 
 This space (in the light-cone gauge) is thus parameterized by the center-of-mass coordinates and the oscillation modes and takes the form $\mathbb{R}^D\times \XX$. 
If we define Cauchy surface and subregions by imposing appropriate conditions on the center-of-mass coordinates $\mathbb{R}^D$ then the above analysis is applicable. 
 \section{Entanglement entropy in open string theory}\label{sec:EEinOSFT}
 In this section we compute entanglement entropy in open SFT using our result in~\cref{eq:EEhalfLI}. This has been 
previously computed by~\cite{Balasubramanian:2018axm} using algebraic methods and canonical quantization. Our approach is based on the replica 
method  which we later generalize to the case of 
closed strings. 
   
 In the light-cone gauge, the quadratic part of the open SFT action takes the form~\cite{Thorn:1988hm}
\be
{\rm I}_{\rm OSFT}= 
\int \sbkt{ DX}_{\rm open}
\ \Phi 
 \bkt{\p_+\p_- + \frac{\pi}{2} \int_0^\pi d\sigma \bkt{-\frac{\delta}{\delta X^I\bkt{\sigma}}\frac{\delta}{\delta X^I\bkt{\sigma}} 
+\frac{1}{4 
\pi^2\alpha'^2} \p_\sigma X^I\bkt{\sigma} \p_\sigma X^I\bkt{\sigma}}}\Phi.
\ee 
 Various terms appearing in this action are to be understood as follows.
  $x^+\equiv 
{X^0+X^{D-1}\over\sqrt{2}}
$ is the light-cone time coordinate\footnote{There are subtleties associated with light-cone quantization such as well-posedness 
of the Cauchy problem, causality, etc. There is a well known method to avoid these subtleties by shifting the coordinates such 
that the constant $x^+$ surfaces are space like~\cite{Burkardt:1995ct}. For the computation of partition functions in Euclidean 
signature such subtleties can be safely ignored. 
}
 and 
 $x^-$ is the zero mode associated with the other linearly independent combination 
\be
{X^0-X^{D-1}\over\sqrt{2}}=x^-+\cdots.
\ee
Here `$\cdots$' denotes $\sigma$-dependent terms which are  determined  in terms of the mode expansion of the transverse 
coordinates $X^I\bkt{\sigma}$ for  $I=1,2,\cdots, D-2$. The transverse coordinates have the mode expansion
\be
X^I\bkt{\sigma}=x^I+\sqrt{2}\sum_{m=1}^{\infty} x_m^I\cos\bkt{m \sigma}.
\ee
The differential operator ${\delta \over \delta X^I\bkt{\sigma}}$ acts on the string-field $\Phi$ as follows
\be
{\delta \over\delta  X^I\bkt{\sigma}}\equiv {1\over \pi}\bkt{{
\p \over \p x^I}+ \sqrt{2}\sum_{m=1}^{\infty}\cos\bkt{m\sigma} {\p \over \p x^I_m} }.
\ee
 The string-field $\Phi$ is defined on the configuration space of all open strings $\MM_{\rm open}$ which is infinite dimensional. 
In the light-cone gauge this space is parameterized by $x^+, x^-$ and $X^I\bkt{\sigma}$ or equivalently
 \be
\MM_{\rm open}= {\mathbb R}^{D}\times\XX,
\ee 
where $\XX$  is an infinite dimensional space parameterized by the coordinates $ x_m^I$.
The measure $\sbkt{D X}_{\rm open}$ is an infinite dimensional integration measure on $\MM_{\rm open}$ which can be written 
in terms of the mode expansion as
\be
\sbkt{D X}_{\rm open} \equiv dx^- dx^+d^{D-2} x\times \prod_{m=1}^\infty d^{D-2} x_m= d^{D} x\times \prod_{m=1}^\infty d^{D-2} x_m.
\ee
 
Using the mode expansion,  the action can be recasted into a more familiar form 
\be
{\rm I}_{\rm OSFT}= \int d^D x \prod_{n=1}^{\infty} d^{D-2} x_m\ \Phi\bkt{\p_+ \p_-
-
\frac12\p_I\p^I 
-\frac12 \sum_{m=1}^{\infty} 
\bkt{
{\p^2\over \p x_m^I \p x_m^I} 
- \bkt{m\over 2 \alpha'}^2 x_m^I x_m^I
}
}
\Phi. 
\ee
 Along ${\mathbb R}^D$ the kinetic operator is the usual Laplacian. Along $\XX$, the kinetic operator is the sum of 
time-independent 
Schrodinger operators for simple harmonic oscillators with frequency ${m\over 2 \alpha'}$ for $m=1,2,\cdots$. 
 
 Let $ \KK_{m}\bkt{t, x_m^I: y_m^I}$ denote the heat kernel for the oscillator part of the kinetic term. Then according 
to~\cref{eq:EEhalfLI}, the entanglement entropy for open strings is 
\be
S_{\rm open}= \frac{  A}{12 \bkt{2\pi}^{D/2-1}}\int_0^\infty {dt\over t^{D/2} }  \prod_{I=1}^{D-2}\prod_{m=1}^{\infty}\Tr 
\KK_{m}\bkt{t, 
x_n^I:x_n'^I}.
\ee
The properly normalized heat kernel for the simple harmonic oscillator can be found in~\cite{zinn-justin:2005}. The trace gives 
the partition function of the simple harmonic oscillator with frequency ${m\over 2\alpha'}$.
\be
\Tr \KK_{m}\bkt{t} = {e^{-{m t \over 4\alpha'}} \over 
1- e^{-{m t \over 2\alpha'}}}.
\ee
After performing the product over $m$ and $I$ and setting $D=26$ we get
\be
 \prod_{I=1}^{D-2}\prod_{m=1}^{\infty}\Tr \KK_{m}\bkt{t}= \eta\bkt{\io t\over 4 \pi \alpha'}^{-24}.
\ee
We scale the integration variable $t\to t\times 4 \pi \alpha'$ to write the entanglement entropy as
\be
S_{\rm open}=
\frac{ A}{12 \bkt{8 \pi^2 \alpha'}^{12}}\int_0^\infty {dt\over t^{13} }  \eta\bkt{\io t}^{-24}.
\ee
This  matches the result of~\cite{Balasubramanian:2018axm} and is also consistent with the effective field theory expectation. It was argued in ~\cite{Balasubramanian:2018axm} 
that their algebraic method does not encounter the `contact terms' of conical entropy which appear in the worldsheet 
computation using the replica method~\cite{He:2014gva}. Here we see that the replica method does not encounter the 
aforementioned contact-terms as well. These contact-terms are related to contributions from the edge-modes. Any analysis 
based on the light-cone gauge fixes the gauge freedom completely. We are essentially studying the dynamics of a collection of 
scalar degrees of freedom and hence do not encounter any contact-terms.
The agreement with~\cite{Balasubramanian:2018axm} is a compelling evidence that the replica method is a reliable tool to compute entanglement entropy in SFT.
\section{Entanglement entropy in closed string theory}\label{sec:EEinCSFT}
We now compute the entanglement entropy in closed string theory and show that it is UV-finite. Let us summarise the result of this section before giving details of the computation. A simple generalization of the open SFT 
action, which includes kinetic term with extra oscillator modes, gives a UV-divergent R\'enyi partition functions and entanglement entropy. This simple generalization does not even give the correct one-loop amplitude for closed 
strings~\cite{Polchinski:1985zf}. The correct one-loop amplitude requires a non-trivial path integral measure~\cite{Sen:1993kb}. After taking the path integral measure into account, we obtain a UV-finite entanglement entropy. 
\subsection{Closed SFT with the canonical kinetic term and path-integral measure}
The development of closed SFT has an exciting and rich history. Light-cone SFT was developed 
in~\cite{Mandelstam:1973jk,Mandelstam:1974hk}. 
The covariant formalism for closed bosonic strings was developed in the seminal work of 
Zwiebach~\cite{Zwiebach:1992ie}(see~\cite{deLacroix:2017lif} for a recent review of covariant formalism for closed 
superstrings). 
Since we are interested in the free light-cone theory, we only need some elementary ingredients which can also be `derived' from the
first quantized theory.  
In the light-cone gauge the action takes the form
\be
{\rm I}_{\rm CSFT}= \int \sbkt{D X}_{\rm closed}
\Phi
\bkt{
\p_+\p_- 
+ \pi 
\int_{0}^{2\pi} d\sigma 
\bkt{
-\frac{\delta}{\delta X^I\bkt{\sigma}}\frac{\delta}{\delta X^I\bkt{\sigma}} +\frac{1}{4 \pi^2\alpha'^2} \p_\sigma X^I\bkt{\sigma} 
\p_\sigma X^I\bkt{\sigma}}
}
\Phi,
\ee
with the mode expansions
\be
\begin{split}
X^I\bkt{\sigma}&=x^I+\sqrt{2}\sum_{m=1}^{\infty}  x_m^I\cos\bkt{m\sigma}
+\tilde{x}_m^I\sin\bkt{ m\sigma},
\\
{\delta \over\delta  X^I\bkt{\sigma}}&= {1\over 2 \pi}\bkt{{
\p \over \p x^I}+ \sqrt{2}\sum_{m=1}^{\infty}\cos\bkt{m\sigma} {\p \over \p x^I_m} 
+\sin\bkt{m\sigma}{\p\over\p \tilde{x}_m^I}
}.
\end{split}
\ee
In addition to the equations of motion resulting from the above action, the closed string-field is also subject to a constraint which 
has its origins in the level matching condition. This can be expressed as
\be\label{eq:const}
\int_0^{2\pi} d\sigma
X'^{I}\bkt{\sigma} {\delta\over \delta X^I\bkt{\sigma}} \Phi=0,
\ee
As before, one can expand the action and constraints in terms of the coordinates $x_m^I$ and $\tilde{x}_m^I$. In particular, the 
constraint takes the following form
\be\label{eq:const2}
\sum_{m=1}^{\infty} m\bkt{x_m^I{\p\over \p \tilde{x}_m^I}-\tilde{x}_m^I {\p \over \p {x}_m^I}} \Phi=0.
\ee
It seems difficult to gain much insight from this form of the constraint. Let's consider an eigenfunction of the kinetic operator 
along the oscillator directions. These are simply the wavefunctions of an infinite collection of simple harmonics oscillators. The 
vacuum satisfies the constraint but it is not obvious which excited wavefunctions also satisfy the constraint.  To explore the 
consequences of the constraint further it is useful to change variables.
The key idea is to note that $X^I\bkt{\sigma}$ and ${\delta \over\delta  X^I\bkt{\sigma}}$ act on the string-field as operators 
which are canonically conjugate to each other, 
\be
\sbkt{X^I\bkt{\sigma}, {\delta \over\delta  X^J\bkt{\sigma'}}}= - \delta^{IJ} \delta\bkt{\sigma-\sigma'}.
\ee
By a functional Fourier transform, one can switch the role of the `coordinate' variable and the `momentum' variable. 
The mode expansion given above provides an infinite set of canonical pairs.
\be
\sbkt{x_n^I, {\p\over \p x_m^J}}=-\delta_{mn}\delta^{IJ},\qquad \sbkt{\tilde{x}_n^I,{\p\over \tilde{x}_m^J}}=-\delta_{mn}\delta^{IJ}.
\ee
The choice is not unique\footnote{This non-uniqueness is related to the ambiguity in identifying canonically conjugate variables for the Poincar\'e conserved charges in string theory. We discuss this in more detail in~\cref{sec:ambiguities}.}. We can equally well assign the `coordinate' interpretation to ${\p\over x_m^I}\equiv {x'}_m^{I}$ with 
$-x_m^I\equiv 
{\p\over \p {x'}_m^{I}}$ being its derivative. We now utilize this freedom to make the following change of variables 
which does not modify the commutation relations given above. 
\begin{alignat}{4}
&x_m^I\to {1\over \sqrt{2}}\bkt{x_m^I+\tilde{x}_m^I}&,
\qquad \qquad
&{\p\over \p x_m^I}\to {1\over \sqrt{2}}
\bkt{{\p\over \p x_m^I}+{\p \over \p \tilde{x}_m^I}}&,\\
&\tilde{x}_m^I\to  {\io \alpha'\over m \sqrt{2}}\bkt{{\p\over \p x_m^I}-{\p\over \p \tilde{x}_m^I}}&,
\qquad\qquad
&{\p \over \p \tilde{x}_m^I}\to {\io m\over \sqrt{2} \alpha'}
\bkt{x_m^I- \tilde{x}_m^I}&.
\end{alignat}
In terms of this new mode expansion, we can write the action as 
\be\label{eq:ICSFTModes}
\begin{split}
{\rm I}_{\rm CSFT}= \int \sbkt{D X}_{\rm closed}
\ &\Phi
\OO \Phi.
\end{split}
\ee
The kinetic operator $\OO$ is given by
\be
\OO={
\p_+\p_- 
- \frac12 \p_I\p^I+\OO_\XX+\OO_{\widetilde{\XX}}},
\ee
where we have defined the operators 
\be
\OO_\XX=\frac12 \sum_{m=1}^{\infty}\bkt{-{\p^2\over \p x_m^I \p x_m^I}
+\bkt{m\over\alpha'}^2
x_m^I x_m^I
},
\qquad 
\OO_{\widetilde{\XX}}
= 
\frac12\sum_{m=1}^{\infty}
\bkt{
-{\p^2\over \p \tilde{x}_m^I\tilde{x}_m^I}
+
\bkt{m\over\alpha'}^2 \tilde{x}_m^I \tilde{x}_m^I
}.
\ee
The constraint~\cref{eq:const} now takes a simpler form
\be\label{eq:constF}
\bkt{\OO_\XX-\OO_{\widetilde{\XX}}} \Phi=0.
\ee
The space of configurations of closed string in the light-cone gauge is
\be
\MM_{\rm closed}= \mathbb{R}^{ D}\times \bkt{\XX\times \widetilde{\XX}},
\ee
where $\XX$ and $\widetilde{\XX}$  are two infinite dimensional spaces encoding the dependence on the oscillator modes 
$x^I_n$ and $\tilde{x}_n^I$. The integration measure on this space is 
\be
\sbkt{D X}_{\rm closed}\equiv d^D x\times \prod_{m=1}^{\infty} d^{D-2} x_m d^{D-2} \widetilde{x}_m
\ee

So far we have insisted on imposing the constraint of~\cref{eq:constF} `by hand'. We can implement the 
constraint~(\ref{eq:constF}) as a consequence of equations of motion\footnote{ In the covariant closed SFT 
requirement of a suitable kinetic term and the inclusion of interaction demands that the string field is subject to a set of 
subsidiary conditions which include $\OO_- \Phi=0$~\cite{Zwiebach:1992ie}. It is not clear to us if these conditions can be relaxed off-shell.} by introducing the  operator
\be
\PP\equiv \int_{-\frac12}^{\frac12} ds\  e^{2\pi \io \alpha' \OO_- s}= {\sin \bkt{\alpha' \pi \OO_-}\over \alpha' \pi \OO_-},
\ee
which projects onto the space of functions which are annihilated by $\OO_-=\OO_\XX-\OO_{\widetilde{\XX}}$. We can now decompose an arbitrary field $\Phi$ as \be
\Phi= \Psi+\varphi,
\ee
such that $\Psi$ is annihilated by $\OO_-$, i.e., $\PP \Psi = \Psi$ and $\varphi$ is not, i.e., $\PP \varphi = 0$. The closed SFT 
action can now be written as 
\be\label{eq:ICSFT22}
{\rm I}_{\rm CSFT}= 
\int \sbkt{D X}_{\rm closed}
\Psi \OO \Psi+ \varphi\varphi,
\ee
 Fields $\varphi$ which are not annihilated by $\OO_-$ are set equal to zero by the equation of motion and only 
the fields $\Psi$ which are annihilated by $\OO_-$ furnish consistent on-shell configurations. Using the projection operator $\PP$ 
and the orthogonality of $\Psi$ and $\varphi$  we can write the above action in terms of the unrestricted fields $\Phi$ as follows
\be
{\rm I}_{\rm CSFT}= 
\int \sbkt{D X}_{\rm closed} \Phi \OO \PP \Phi+\Phi \bkt{\mathbf{1}-\PP} \Phi.
\ee
Upon expanding the above quadratic operator there will be cross terms between $\Psi$ and $\varphi$ but they integrate to zero 
because of the orthogonality. 
The operator that is appearing in the above action is actually equal to the operator $\OO^\PP$. To see this note that we can 
rewrite $\OO^\PP=\exp\bkt{\PP\log \OO}$. Since $\PP$ and $\OO$ commute, this rewriting is unambiguous. Next we expand the 
exponential in powers of $\PP\log\OO$
\be
\exp\bkt{\PP \log \OO}=\sum_{n=0}^{\infty}{ \PP^n\bkt{\log \OO}^n\over n!} = \mathbf{1}+\sum_{n=1}^\infty\PP {\bkt{\log 
\OO}^n\over 
n!}=\mathbf{1}+\PP\bkt{\exp\bkt{\log \OO}-\mathbf{1}}= \OO\PP +\bkt{\mathbf{1}-\PP}.
\ee
So we finally arrive at the action 
\be\label{eq:ICSFT2}
{\rm I}_{\rm CSFT}= 
\int \sbkt{D X}_{\rm closed}  \Phi \OO^\PP \Phi.
\ee

Given this action, the computation of entanglement entropy is a straightforward generalization of the open string case. 
On the $n$-fold branched cover of $\MM_{\rm closed}$ the heat kernel factorizes. After computing the trace over the `zero-mode' 
directions the entanglement entropy takes the form 
\be
S_{\rm closed}
= { A\over 12 \bkt{2\pi }^{D/2-1}} \int {dt\over t^{D/2}} \Tr \PP \KK_{\OO_\XX+\OO_{\widetilde{\XX}}}
\ee
Expressing the heat kernel in terms of the operators and after scaling $t$ appropriately the above expression becomes
\be
S_{\rm closed}={A\over 12 \bkt{4\pi^2 \alpha'}^{12}}\int {dt \over t^{D/2}} \int_{-\frac12}^{\frac12} ds
\Tr  q^{\alpha' \OO_\XX} \bar{q}^{\alpha' \OO_{\widetilde{\XX}}},
\ee
where $q=e^{2\pi i \bkt{s+\io t}}\equiv e^{2\pi \io \tau}$.  The trace over the oscillators gives familiar factors of Dedekind eta 
function 
\be\label{eq:COtrace}
 \Tr  q^{\alpha' \OO_\XX} \bar{q}^{\alpha' \OO_{\widetilde{\XX}}}=\Big|\eta\bkt{\tau} \Big|^{-48}.
\ee
Finally we write the integral in terms of the complex variable $\tau$ to get
\be
S_{\rm closed}=
\frac{ A}{24 \bkt{4 \pi^2 \alpha'}^{12}}\int_{\SS} {d\tau d\bar{\tau}\over \bkt{\Im\tau}^{13} }  \Big|\eta\bkt{\tau}\Big|^{-48},
\ee 
where the integral is over the strip $\SS=\big\{
\tau | \Im \tau \geq 0,  -\frac12\leq\Re \tau< \frac12
\big\}$. This resembles the torus amplitude in closed strings. However, if we interpret $\tau$ as the modulus of the torus, the integration region is not restricted to the fundamental 
domain.  The entanglement entropy computed above is divergent. In 
the $\Im \tau\to \infty$ region of the integration, it has divergences due to the closed string tachyon but this is not interesting. We 
expect such a divergence to go away in a consistent string theory. It also has the usual UV-divergence of quantum field theory in 
the $\Im \tau\to 0$ limit of integration region. In fact, the result can be written as a sum over entanglement entropies of different 
fields in the closed string spectrum. In that sense, one may argue that this is indeed the expected result and hence also has 
usual UV-divergences. 

But we stress that the physics of closed string theory, even in the free limit, is different than that of a collection of free 
fields in the closed string spectrum. The most lucid example of this fact is the UV-finiteness of one-loop vacuum amplitude in 
closed string theory. The vacuum amplitude is not just a sum over individual, UV-divergent vacuum amplitudes of all fields in the closed string 
spectrum. 
Path integral with a trivial integration measure and the canonical kinetic term that appeared in the closed SFT action above, does not  give the correct vacuum amplitude for 
closed string theory. We discuss this in more detail in the next section. 
\subsection{Path integral measure for closed SFT}\label{subsec:nckt}
The vacuum amplitude from the closed SFT action~(\ref{eq:ICSFT2}) with a trivial path integral measure is given by
\be\label{eq:divve}
\begin{split}
\VV &= -\log \ZZ=\frac12 \Tr \log \OO^\PP, \\
&= 
-\frac14 {V \over \bkt{4\pi^2 \alpha'}^{13}} \int_{\SS}{d\tau d\bar{\tau} \over \bkt{\Im \tau}^{14}} \Big|\eta\bkt{\tau}\Big|^{-48}.
\end{split}
\ee
This is not the same as the one obtained from the worldsheet computation~\cite{Polchinski:1985zf}. The difference is precisely 
that of the integration region in the complex $\tau$-plane. The world sheet computation has an integration over the fundamental domain ${\cal F}$ of the 
torus and hence gives a finite (up to tachyonic divergences) result. If we interpret $\tau$ as the modulus of the torus then the 
strip $\SS$ covers the moduli space of torus an infinite number of times and this over counting leads to an infinite answer. 

In 
ordinary quantum field theory, vacuum amplitude only gives an overall phase and hence drops 
out in all expectation values. However, when one considers the coupling to gravity, the one-loop vacuum amplitude sources the 
geometry and hence it is an observable in a theory of quantum gravity. Therefore, we must ensure that our treatment of the closed SFT path integral reproduces the correct one-loop vacuum amplitude. 
Zwiebach and Sen addressed the subtleties regarding vacuum amplitude in closed SFT in~\cite{Sen:1993kb}. In the following we briefly summarize their relevant results. 

 The one-loop vacuum amplitude in closed SFT receives contributions from three terms
\be
\VV_{\rm w.s}=\frac12 \Tr \log \OO_K+ {\rm I}_0+ \frac12 \log \rho,
\ee
where $\OO_K$ is the operator in the quadratic part of the action. 
 ${\rm I}_0$ is a \emph{constant} obtained by integrating the world-sheet CFT partition function over a subset of the moduli space of the torus. The last term is the contribution from the so-called `string-field measure', i.e., one defines the partition function with the following measure over the string-fields $\Psi_i$
\be
\sqrt{\rho} \int \prod_i d\Psi_i \ \exp\bkt{-{\rm I}_{\rm CSFT}}.
\ee
 ${\rm I}_0$ and  $\frac12 \log \rho$ enter the path integral in the same way and we do not need to consider these separately therefore we set ${\rm I}_0=0$ henceforth. The measure $\rho$ and the operator $\OO_K$ are not uniquely determined individually. Field redefinitions, in general, will change the operator $\OO_K$ as well as $\rho$ but not the combination $\Tr \bkt{\OO_K}+\log\rho$. This ensures that the one-loop vacuum amplitude is a physical observable which is independent of the choice of the off-shell data such as field redefinitions.  

Let us now use these insights to obtain a proper integration measure and kinetic term for the light-cone SFT. 
We must use a non-trivial path integral measure to obtain the correct one-loop amplitude. Equivalently, we can use the trivial measure but a different field basis
\footnote{\label{fn:csbcomment}In the covariant closed SFT one can obtain a finite contribution to the vacuum amplitude from 
a modified kinetic term. The finite contribution involves integration of the worldsheet conformal field theory partition function over 
a strip in the moduli space of the torus. It seems necessary to add a constant term in the action to get the correct vacuum 
amplitude which involves integration over the fundamental domain of the torus. We thank Barton Zwiebach for discussion on this 
point.}. We now find a field basis $\Phi'$ for which the flat integration measure, i.e., $\rho'=1$ gives the correct one-loop vacuum amplitude. 
Since we are limiting ourselves to the free closed SFT, we look for field redefinitions which do not introduce any cubic or 
higher-order 
terms in fields. The eventual consequence of such a field redefinition is to change the kinetic term so 
that
\be\label{eq:IcsftMod}
{\rm I}_{\rm CSFT}= \int \sbkt{DX} \Phi' \OO^\PP\times f\bkt{\OO,\OO_-} \Phi'.
\ee
Here $f\bkt{\OO,\OO_-}$ is a differential operator which  \emph{must} be chosen so that
\begin{enumerate}
\item The on-shell content of the theory does not change. This can be guaranteed by showing that $\OO^\PP$ and $\OO^\PP 
f\bkt{\OO,\OO_-}$ have the same kernel. 
\item The inverse of the kinetic operator only has simple poles at $\OO^\PP=0$, i.e., at on-shell states. This can be ensured by 
arguing that $f$ does not vanish on-shell. 
\item The vacuum amplitude from the modified action coincides with the worldsheet computation.  
\end{enumerate}

Note that these three requirements do not determine the kinetic operator uniquely.  One can modify it further without spoiling the 
above conditions. We emphasize, however, that this ambiguity does not lead to an ambiguity in the computation of entanglement entropy. To see this more explicitly, consider two operators  $\OO^\PP f\bkt{\OO,\OO_-}$ and $\OO^\PP f\bkt{\OO,\OO_-} g\bkt{\OO,\OO_-}$ which satisfy the above three requirements. Since the vacuum amplitude resulting from the both must be the same this implies
\be
\Tr \log g\bkt{\OO,\OO_-}=0.
\ee
It then follows that such modification would not  change the result of the path integral over the replicated manifold and would not 
affect the result of the entanglement entropy.

We now find the function $f\bkt{\OO,\OO_-}$ so that the third requirement is satisfied by construction. We will then demonstrate 
that our choice also satisfies the first two requirements. 
The vacuum amplitude as computed from worldsheet methods takes the following form
\be
\VV_{\rm w.s}=-\frac14 {V\over \bkt{4\pi^2 \alpha'}^{13}} \int_{{\cal F}}{d\tau d\bar{\tau} \over \bkt{\Im \tau}^{14}} 
\Big|\eta\bkt{\tau}\Big|^{-48}
\ee
Using $\tau=s+\io t$ the above expression can be written in terms of a trace involving operators $\OO$ and $\OO_-$. Note that
\be
\Tr e^{2\pi\io \alpha' s \OO_-} e^{-2\pi\alpha' t \OO} =\Tr q^{\alpha' \OO_\XX} \bar{q}^{\alpha' \OO_{\widetilde{\XX}}} e^{+\pi\alpha' 
t \p_\mu\p^\mu}. 
\ee
The trace over the oscillators, i.e., the first two factors is given in~\cref{eq:COtrace} and the trace of the last factor is computed 
in~\cref{eq:trko}. We get
\be
\Tr e^{2\pi\io \alpha' s \OO_-} e^{-2\pi\alpha' t \OO} = \Big|\eta\bkt{\tau}\Big|^{-48}\times {V\over \bkt{4\pi^2\alpha' t}^{13}},
\ee
We can now write the worldsheet vacuum amplitude as
\be\label{eq:wsTrace}
\begin{split}
\VV_{\rm w.s}
&=-\frac12 \Tr \int_{-\frac12}^{\frac12} ds 
e^{2\pi\io \alpha' s \OO_-}
\int_{\sqrt{1-s^2}}^{\infty} {dt\over t}
 e^{-2\pi\alpha' t \OO} , 
\\
&=-
\frac12\Tr \int_{-\frac12}^{\frac12} ds \ e^{2\pi\io \alpha' s \OO_-} E_1\bkt{2\pi\alpha'\sqrt{1-s^2} \OO},
\end{split}
\ee
where $E_1\bkt{x}$ is the exponential integral defined (for values of $x$ off-the negative real axis) as~\cite{2007xxxi}
\be\label{eq:expInt}
E_1\bkt{x}
=
\int_{1}^{\infty} {e^{-t x}\over t} dt =
-\gamma-\log x-\sum_{n=1}^{\infty} {\bkt{-x}^n\over n n!}.
\ee
Comparing the last line of~\cref{eq:wsTrace} with the vacuum amplitude $\frac12\Tr\log \OO^\PP f\bkt{\OO,\OO_-}$ obtained 
from the closed SFT action in~\cref{eq:ICSFTModes} gives\footnote{It is instructive to 
compare this kinetic operator with the one obtain by Zwiebach and Sen in sec.~(6.3) of~\cite{Sen:1993kb}. Their kinetic operator 
can be written as $\exp\bkt{-E_1\bkt{2 \alpha' a \OO}}$. Upon computing the trace this gives the integral of the worldsheet 
partition function over a strip in the moduli space of the torus as mentioned in~\cref{fn:csbcomment}. Geometrically this 
corresponds to including all tori built by joining opposite ends of a cylinder of length greater than or equal to $2a$.}
\be\label{eq:modKO}
\OO^\PP f= \exp\bkt{
-\int_{-\frac12}^{\frac12} ds e^{2\pi \io\alpha' s \OO_-}
E_1\bkt{2\pi\alpha' \sqrt{1-s^2} \OO}
}.
\ee
Using the series representation of the exponential integral~\cref{eq:expInt} we can also find a power series expansion for $\log f$ 
\be\label{eq:log}
\log f= 
\PP \log\bkt{2\pi \alpha' e^\gamma}
+\int_{-\frac12}^{\frac12}
ds e^{2\pi \io s\OO_-}
\sbkt{\log\sqrt{1-s^2}
+\sum_{n=1}^{\infty}
{\bkt{-2\pi\alpha' \sqrt{1-s^2} \OO}^n \over n n!}
}.
\ee  
For any finite value of $\OO$ the function $\log f$ is finite. The modified kinetic operator is highly non-local as it involves infinite number of derivatives. Nevertheless, it gives the correct vacuum amplitude and satisfies the other two properties described after~\cref{eq:IcsftMod} as we prove below.

 First we show that  $\OO^{\PP}$ and $\OO^{\PP} f$ have the same kernel. If $\PP=1$ and $\OO=0$ then $f$ is a finite quantity and 
hence $\OO^\PP f=0$. To prove the converse let's assume $\OO^\PP f = 0$ but $\OO^\PP\neq 0$. This 
can happen in two ways: (1). $\PP=0$. It is then clear that $\log f$ is finite for any finite value of $\OO$ and hence $\OO^\PP 
f\neq 0$ leading to a contradiction. (2). $\PP=1$  but $\OO\neq 0$. In this case $\OO^\PP f= \OO f $. Now, again for any finite 
value of $\OO$, $f$  is a finite quantity so $\OO$ must vanish, satisfying the first requirement.

  The second requirement is also satisfied because $f$ is a finite quantity as $\OO\to 0$ so ${1\over \OO f}$ only has simple 
poles at $\OO=0$.  

Let us now go back to the field basis with canonical kinetic term to determine the non-trivial path integral measure. The action has a canonical kinetic term in the basis $\Phi$ which are related to $\Phi'$ as
\be
\Phi= f^{\frac12} \Phi'.
\ee
The path integral measure is 
\be
D\Phi'= \det |f^{-\frac12}|  D\Phi.
\ee
So that the action with canonical kinetic term gives the correct one-loop vacuum amplitude if one uses the non-trivial measure $\rho=\det|f|$. In summary, to obtain the correct vacuum amplitude from the SFT action one must carefully specify the integration measure in the path integral. The choice of integration measure depends on the field basis. In particular, in the field basis with flat integration measure the kinetic term is not canonical and in the field basis with the canonical kinetic term there is a non-trivial integration measure. 

Note that the integration measure we have obtained above is field independent therefore in computations of Green's functions it drops out. Since different kinetic terms give different propagators we obtain different Green's functions in the field basis $\Phi$ and $\Phi'$. The S-matrix elements, however, are the same regardless of the choice of the field basis.  This follows because the field redefinition relating $\Phi$ and $\Phi'$ satisfies the usual criteria for the S-matrix equivalence, i.e., the field $\Phi'$ has non-zero matrix element between the vacuum and the one-particle state for field $\Phi$ and vice versa. This in turn follows because on-shell $f\neq 0$. 
\subsection{Entanglement entropy with the correct path integral measure}
To compute the R\'enyi partition functions We need the representation of $\log 
\OO^\PP f$ on the $n$-fold branch cover of $\MM_{\rm closed}$. As before, the trace factorizes into contributions from 
oscillators and zero-modes 
\be
\begin{split}
\log \ZZ&=-\frac12 \Tr \log \OO^\PP f
=
\frac12 \Tr \int_{-\frac12}^{\frac12} ds e^{2\pi \alpha'  s \OO_-}
\int_{\sqrt{1-s^2}}^\infty {dt\over t} e^{-2\pi\alpha' t \OO},\\
&= \frac12 \int_{-\frac12}^{\frac12} ds \int _{\sqrt{1-s^2}}^\infty {dt\over t}  \Tr_{\XX} \bkt{q^{\alpha' \OO_\XX}} 
\Tr_{\widetilde{\XX}} \bkt{q^{\alpha' \OO_\XX}} 
\Tr_{\mathbb{R}^{26}} e^{+\pi t\alpha' \p_\mu \p^\mu}. 
\end{split}
\ee
To compute the trace over the branched cover one only needs to replace the zero-mode factor, i.e., $\mathbb{R}^{26}$ by the 
appropriate branched cover.  The trace of
$ e^{-t \OO\bkt{\p_\mu\p^\mu}}$ on that branched cover is computed in the appendix (see~\cref{eq:hkp2}). The trace over the 
oscillators gives well known factors of Dedekind eta function. The logarithm of the R\'enyi partition function is
\be
\log\ZZ\bkt{n}={A\over 48 \bkt{4\pi^2 \alpha'}^{12}} {1-n^2\over n}\int_{\cal F}{d\tau d\bar{\tau}\over\bkt{ \Im \tau}^{13}}\Big| 
\eta\bkt{\tau}\Big|^{-48}+n \log \ZZ.
\ee
The entanglement entropy computed from this is 
\be
S\label{eq:SClosedFinal}
=
{A\over 24\bkt{4\pi^2\alpha'}^{12}} \int_{{\cal F}} {d\tau d\bar{\tau}\over \bkt{\Im \tau}^{13}} \Big|\eta\bkt{\tau}\Big|^{-48},
\ee
which is UV-finite. There are tachyonic divergences but these will be absent in the superstrings.  

If one interprets the parameter $\tau$ as modulus of a torus then the result for entanglement entropy is \emph{not} 
modular invariant. Here we use the term modular invariance in the same sense in which the vacuum amplitude of closed strings 
is modular invariant:  the answer can be written as an integral of a modular invariant function over the fundamental domain of the 
torus with a modular invariant integration measure\footnote{ In two dimensional conformal field theories the entanglement entropy also fails to be modular invariant~\cite{Lokhande:2015zma}.}. The lack of modular invariance in our result is not entirely surprising.
R\'enyi partition function is analogous to the partition function of a sigma model with branched cover target space. 
As explained in the introduction such a sigma model is not a conformal field theory and this may explain the lack of modular invariance in our result. 

Nevertheless, the lack of modular invariance means that the above result could not be the complete answer for entanglement entropy in closed strings. In fact there is a good reason to believe that there are extra contributions to the entanglement entropy of closed strings that we have not captured. Closed SFT has a gauge symmetry and our computation in the light-cone gauge misses the contribution from edge-modes, i.e., gauge degrees of freedom that become dynamical at the entangling surface and contribute non-trivially to the entanglement entropy. It is useful to compare the situation with four-dimensional Abelian gauge theory. In the light-cone gauge the theory has two degrees of freedom and the entanglement entropy is just twice the entanglement entropy of a free scalar. But this differs from the complete answer that one obtains by considering the gauge theory on the branched cover spacetime prior to gauge fixing~\cite{Kabat:1994vj}. The difference is attributed to the edge-modes\cite{Donnelly:2014fua,Donnelly:2015hxa,Donnelly:2016auv}.   
We believe that the full answer for entanglement entropy in closed SFT, after a careful analysis of contributions from {edge-modes} would be modular invariant.

\section{Discussion, caveats and outlook}\label{sec:conclusion}
In this paper we have taken initial steps towards understanding entanglement entropy in closed string theory using the formalism 
of SFT. We only needed some elementary ingredients from closed light-cone SFT and we demonstrated that the resulting 
entanglement entropy is UV-finite. The mechanism responsible for the finiteness of the entanglement entropy is the same that makes the one-loop 
vacuum amplitude UV-finite. While this is encouraging progress, we have taken a pragmatic approach so far, neglecting subtleties associated with defining subregions and entanglement entropy in string theory as well as the algebraic aspects of finiteness of entanglement entropy. In the rest of this paper we discuss these issues in light of our computation.
\subsection{Ambiguities in defining spacetime and subregions}\label{sec:ambiguities}
There are conceptual issues related to entanglement and the definition of subregions in a theory of gravity. In a theory with a gauge symmetry degrees of freedom in two regions cannot be factorized unambigously. The extended phase space approach of~\cite{Donnelly:2016auv,Speranza:2017gxd} addresses this issue by introducing extra degrees of freedom at the entangling surface, the so called \emph{edge-modes}, which contribute non-trivially to the entanglement entropy. The analysis of~\cite{Balasubramanian:2018axm} for the case of open strings suggest that a similar 
picture should hold in closed string theory. The covariant SFT (at least in the $g_s\to 0$ limit) can be used to shed some light on the extended phase space of closed strings. While the issues related to gauge invariance and edge-modes present challenging technical problems there are inherently stringy ambiguities in defining the spacetime and subregions. 

Let us discuss two important choices that we had to make in order to define and compute entanglement entropy.  Even though the detailed expression for entanglement entropy certainly depends on these choices, we 
believe that the general structure, i.e., the finiteness of entanglement entropy is independent of them. 
 First of all we had to make a choice of \emph{physical} time  to define a Cauchy surface in the space of strings and a 
component of space to define a subregion. In the light-cone gauge, these choices may seem natural but in string theory there is 
no unambiguous way to make these choices.
String theory has a well-defined notion of momenta $p^\mu$ as the conserved charges associated with translations in 
$X^\mu\bkt{\sigma}$. There is, however, no unique way to choose the coordinate $x^\mu$ to label the spacetime 
$\mathbb{R}^{D}$. 
One way is to impose the canonical commutation relation
\be\label{eq:xpcomm}
\sbkt{x^\mu,p^\nu}=\io \eta^{\mu\nu}.
\ee
Let us define $x^\mu$ by integrating $X^\mu\bkt{\sigma}$ over the length of the string against a projection $\Pi^{\mu}{}_{\nu}$ 
such that
\be\label{eq:xdef}
x^\mu=\int_0^{2\pi}  d\sigma \Pi^\mu{}_\nu X^\nu\bkt{\sigma},\qquad \int_0^{2\pi} d\sigma \Pi^\mu{}_\nu=\delta^\mu{}_\nu.
\ee
All such $x^\mu$ satisfy the canonical commutation relation of \cref{eq:xpcomm}. The definition~(\ref{eq:xdef}) picks out the 
center of the mass of the string if we choose $\Pi_{\mu}{}^{\nu}={1\over 2\pi} \delta_{\mu}{}^\nu$. But it is 
not the only choice. One can easily arrange to have $x^\mu$ equal to a linear combination of oscillator modes of various 
frequencies. It is also possible to have different choices along different directions. For example, $x^0$ given by the center of 
mass of $X^0\bkt{\sigma}$ and $x^1$ given by an oscillator mode of $X^1\bkt{\sigma}$. If we make such a choice, then we do not 
have the $\U{1}$ symmetry in the $\bkt{x^0,x^1}$-plane which was necessary in our computation of heat kernels on the 
branched cover. 

Second key choice that we made in our analysis was the factorization of the configuration space of strings. Our choice, based on the center-of-mass coordinate, allowed for a tractable computation and brought out interesting aspects. It is desirable to have a factorization based on the position of the entire string. The subtle aspect here is to understand how to factorize the string which crosses the boundary. 

 The ambiguities in defining spacetime and subregions echo such subtleties familiar in diffeomorphism and gauge invariant theories. A resolution of these subtleties in string theory is needed to better understand entanglement entropy in string theory. Unfortunately, our analysis does not resolve these issues. It is, nevertheless, clear that a factorization based on the center-of-mass can be used a convenient starting point in the studies of entanglement properties in string theory.  
\subsection{Non-locality and the algebraic structure}
Another important conceptual question is what does the finiteness of entanglement entropy teach us about the algebraic structure of quantum gravity? Entanglement 
entropy in our computation is rendered finite by non-local nature of  closed strings. Since entanglement entropy is a property of the algebra of observables of the quantum system this suggests that the underlying algebraic structure in quantum gravity is different than that of a local quantum field theory. For perturbative quantum gravity, such non-locality and its consequences for the underlying algebraic structure are discussed 
in~\cite{Giddings:2015lla,Donnelly:2015hta,Donnelly:2016rvo}. 
It was argued in these papers that diffeomorphism invariant observables  have non-local commutators which fail to vanish at spacelike separations. The non-locality appears at first order in the Newton's constant. Analogous results can be obtained in string theory using the framework of SFT~\cite{Lowe:1995ac}. The commutator of string fields fail to vanish at space-like separations at first order in the string-coupling. This might seem counterintuitive given the fact that the scale of non-locality in string theory is controlled by the string length $\sqrt{\alpha'}$ and not the string coupling. Moreover, our result for one-loop entanglement entropy is independent of the string-coupling but it depends on $\alpha'$. But a careful analysis show that the failure of spacelike commutators to vanish is due to the \emph{non-local} interactions between string-fields. The observables that were considered by~\cite{Lowe:1995ac} and~\cite{Giddings:2015lla,Donnelly:2015hta,Donnelly:2016rvo} only see the non-locality when interactions are involved. Entanglement entropy and vacuum amplitude can probe locality even when interactions are absent. %
\subsection{Future directions}
The objective of this paper was to show that the entanglement entropy in closed string theory is UV-finite and it can be computed 
using the framework of string field theory.  We briefly commented on subtleties in defining spacetime and subregions in string theory. There are many directions in which our analysis can be extended to gain a better understanding. We close this paper 
by mentioning a few avenues for further study.
\subsubsection{Entanglement entropy in superstring theory}
A natural extension of this work is to compute the entanglement entropy in superstring theory. Field theory of superstrings have 
been formulated relatively recently (see~\cite{deLacroix:2017lif} for a review) and has already lead to interesting applications. As 
we observed for the case of bosonic strings, the entanglement entropy involved the oscillator partition function as a term in the 
integrand. If the same pattern persists for the case of superstrings entanglement entropy would be zero. This has been argued 
in~\cite{Mertens:2016tqv}. The vanishing of the vacuum amplitude in superstrings is a result of the 
target space supersymmetry and the computation of entanglement entropy via the replica method breaks all spacetime 
supersymmetries. Moreover, divergences in entanglement entropy renormalize the gravitational coupling 
constant and loops of both fermions and bosons contribute to this renormalization with the same sign~\cite{Larsen:1995ax}. 
Therefore, it will be very interesting to study the entanglement entropy for superstrings and see which of the points of views gets 
validated. If entanglement entropy indeed vanishes for superstrings then it would be interesting to identify the source of the necessary negative contributions. Generally these contributions are attributed to contact terms and edge-modes~\cite{Donnelly:2012st} but in a gauge fixed computation these are absent. 
\subsubsection{The algebraic method}
It would also be interesting to compute the entanglement entropy in closed string field theory using the algebraic method  and 
canonical quantization.
This requires identifying a set of canonical pairs on some Cauchy slice in the configuration space of closed strings and then 
imposing canonical commutation relations. The non-canonical kinetic term presents a major obstacle in this regard---one needs to define and infinite number of canonical momenta. We believe that this infinitude then conspires to give a UV-finite result.
\subsubsection{Interactions}
Another interesting direction is to include interactions. Although the full action of the closed SFT has an infinite number of terms\footnote{
In fact this is only true for covariant SFT. The light-cone SFT is cubic as was shown in~\cite{Giddings:1986rf
}. See~\cite{Baba:2009ns,Ishibashi:2011pc,Ishibashi:2016mek,Ishibashi:2013nma,Ishibashi:2017xpn,Ishibashi:2018avw} for recent progress on higher-loop amplitudes in the light-cone SFT. We thank a referee for emphasizing this point and brining these references to our attention.
},
important insights can be obtained by just including the fundamental three string vertex. Within the replica method frame of work, 
the correction due to interactions boils down to computing amplitudes on the branched cover~\cite{Hertzberg:2012mn}. Since the 
branched cover is not an on-shell string background, the definition of off-shell amplitudes in string theory~\cite{Sen:2014pia} is 
expected to play a role here. Another approach will be to use the Susskind-Uglum prescription~\cite{Susskind:1994sm} for computing the off-shell generating functional at higher genus.  
\subsubsection{Covariant SFT}
It will be very instructive to compute the
entanglement entropy using covariant SFT. This will circumvent issues 
associated with the light-cone quantization. This would be the natural framework to address issues related to the  edge 
modes.
\subsubsection{Worldsheet perspective}
 Another important conceptual point is to make contact with the worldsheet description of entanglement entropy. It would be 
interesting to see if our result can be understood in terms of Susskind and Uglum's prescription to compute string theory partition 
function on off-shell backgrounds~\cite{Susskind:1994sm}. Another intriguing proposal in this regard is~\cite{Mazenc:2019ety} 
which introduces the idea of target space entanglement entropy which seems a natural notion to study entanglement entropy in 
string theory from a worldsheet point of view. There, however, are various subtleties regarding reparameterization invariant first quantized systems. Perhaps a modest goal is to better understand the definition and computation of entanglement entropy for relativistic point particle.
 
 We hope that this work will lead to further study in some of the above issues and a better understanding of entanglement 
entropy in closed string theory. 
 \section*{Acknowledgements}
It is a pleasure to thank A. A. Ardehali, A. Dabholkar, G. Festuccia, R. Mahajan, and O. Parrikar for valuable comments and suggestions 
 on this work. I also acknowledge valuable feedback and detailed comments by J. Minahan and B. Zwiebach on a manuscript.
 This work is supported by the Knut and Alice Wallenberg Foundation,
Stockholm, Sweden.
\appendix
\addtocontents{toc}{\protect\setcounter{tocdepth}{0}}
\section{Trace of the heat kernel on the n-fold branched cover}\label{app:ancont}
In this appendix we provide details regarding the computation of the trace of Lorentz-invariant heat kernel on the n-fold branched 
cover. In this case the theory has $U\bkt{1}$ symmetry in the $\bkt{x^1,x^2}$-plane and the heat kernel on $\MM_n$ can be 
written in terms of the heat kernel on $\MM$ as in~\cref{eq:hkonMq}.

For a Lorentz invariant theory $\mathcal{O}=\OO\bkt{\p_\mu\p^\mu}$. By inserting $1=\int {d^D p\over \bkt{2\pi}^D} 
\ket{p}\bra{p}$ 
in the definition of the heat kernel we get the momentum space expression
\be\label{eq:KKinP}
\KK_\OO\bkt{t,x:x'}\ = \int \frac{d^dp}{\bkt{2\pi}^d} 
e^{\io p.\bkt{x-x'}} e^{-t\OO\bkt{-p^2}}
\ee 
Trace of the above quantity is
\be\label{eq:trko}
\Tr \KK_{\OO}\bkt{t}\ =\ \int d^D x \KK_\OO\bkt{t,x:x}=  V \Omega_{D-1} \int_0^\infty \frac{dp}{\bkt{2\pi}^D}\,  p^{D-1} 
e^{-t\OO\bkt{-p^2}}={V 
\over \bkt{2 \pi t}^{\tfrac D2}},
\ee
where $V$ is the volume of the spacetime and $\Omega_{n}={2\pi^{n+1\over 2} \over \Gamma\bkt{n+1\over 2}}$ is the surface 
area of $n$-sphere with unit radius. The first term that appears in the heat kernel on $\MM_n$ as given in \cref{eq:hkonMq} is  
$\KK_\OO\bkt{t,x:x'}$. 
The trace of this term on $\MM_n$ is just $n$ times the expression derived above in \cref{eq:trko}. The 
second term in $\KK_\OO^{(n)}$ is
\be\label{eq:2ndT1}
 T_2=\frac{1}{4\pi \io n}\int_{\CC} dz
\cot\bkt{\frac{z}{2 n}} \KK_\OO\bkt{t,\phi-\phi'+z},
\ee
where the dependence on rest of the coordinates is suppressed and only the polar angle is shown. To find the trace of the 
second term we start by writing~\cref{eq:KKinP} using spherical coordinates in momentum space. We set up the coordinates so 
that the angle between $p^\mu$ and $\bkt{x-x'}^\mu$ is $\theta$. Then the dependence on the rest of $D-2$ angular coordinates 
drops out and integration over those simply give a factor of $\Omega_{D-2}$. We have
\be
\KK_{\OO}\bkt{t,x:x'}=
{\Omega_{D-2}\over \bkt{2\pi}^D}\int_{0}^{\infty} dp \int_0^\pi d\theta p^{D-1} \sin^{D-2}\theta e^{\io p |x-x'| \cos\theta} 
e^{-t\OO\bkt{-p^2}}
\ee
Using 
polar coordinates in the $\bkt{x_1,x_2}$-plane we can write
\be
|x-x'|^2= r^2+r'^2-2r r'\cos\bkt{\phi-\phi'}  + |x_{\perp}-x_{\perp}'|^2, 
\ee
where $x_\perp$ denotes the $D-2$ coordinates on the transverse space. In computing the trace of the term~(\ref{eq:2ndT1})  
we set all coordinates equal and then integrate over $\MM_n$. The integral over the transverse coordinates simply give the area 
$A$ of the entangling surface. Integral over the polar angle $\phi$ gives a factor of $2\pi n$ and we get
\be
\Tr T_2
=
{ n A \Omega _{D-2}\over \bkt{2\pi}^{D-1}} 
 \int_{\CC}\
\frac{dz}{4\pi \io n } \cot\bkt{\frac{z}{2 n}}
 \int_0^{\infty} dr\  r
\int_{0}^\infty dp \ \int_{0}^\pi d\theta \, p^{D-1} 
e^{-t\OO\bkt{-p^2}} 
e^{ 2 i r p \sin\bkt{\frac{z}{2}}\cos\theta}\sin^{D-2}\theta
\ee 
We perform the integral over $\theta$ using
\be
\int_{0}^\pi d\theta \sin^{n}\theta e^{i a \cos\theta}=
\sqrt{\pi} \bkt{\frac2a}^{\tfrac{n}{2}} \Gamma\bkt{\frac{n+1}{2}} J_{\frac{n}{2}}\bkt{a}
\ee
and for the $r$-integration, we change the variable from $r$ to $a$ by setting $r={a\over 2 p \sin \bkt{z\over 2}}$ so that
\be
\Tr T_2= { n A \Omega_{D-2} \sqrt{\pi} \Gamma\bkt{\frac{D-1}{2}} 2^{\tfrac D2-3}
\over \bkt{2\pi}^{D-1}
}
 \int_{\CC}\
\frac{dz}{4\pi \io n } {\cot\bkt{\frac{z}{2 n}}\over \sin^2\bkt{z\over 2}}
\ 
 \int_{0}^\infty dp  \, p^{D-3}
e^{-t\OO\bkt{-p^2}}
\int_0^{\infty} da\ a^{2-\tfrac D2} J_{\tfrac D2-1}\bkt{a} 
\ee
Now perform the $a$-integral and the contour integral using 
\be
\int_0^\infty da\  a^{1-n} J_{n}\bkt{a}= {2^{1-n}\over \Gamma\bkt{n}},\qquad 
 \int_{\CC}
 \frac{dz}{4\pi \io n}
 \frac{\cot\bkt{\tfrac{z}{2 n}}}{\sin^{2}\bkt{\tfrac z2}}
 =
 \frac{1}{3 n^2}\bkt{1-n^2}
\ee
and use the value of $\Omega_{D-2}$ to get
\be
\Tr T_2= { A  \pi^{\tfrac D2}
\over 3 \bkt{2\pi}^{D-1} \Gamma\bkt{\tfrac D2-1}
} \bkt{1-n^2\over n^2}
\ 
 \int_{0}^\infty dp  \, p^{D-3}
e^{-t\OO\bkt{-p^2}}.
\ee
Combining this with the first term, we finally obtain the expression for the trace of heat kernel on $\MM_n$
\be\label{eq:FinalMqTrace}
\Tr \KK_\OO^{\bkt{n}}\bkt{t}
= {2 n V\over \bkt{4\pi}^{\tfrac D2} \Gamma\bkt{\tfrac D2}}
\int dp\, p^{D-1} e^{-t \OO\bkt{-p^2}}
+
{
  A\over 6\bkt{4\pi}^{\tfrac D2-1}
 \Gamma\bkt{\tfrac D2-1}}
 {1-n^2 \over n}
\int dp\, p^{D-3} e^{-t \OO\bkt{-p^2}}. 
\ee
For $\OO\bkt{-p^2}={p^2\over 2}$ this becomes
\be\label{eq:hkp2}
\Tr \KK_{p^2\over 2}^{\bkt{n}} \bkt{t}
=
{n V \over \bkt{2 \pi t}^{\tfrac D2}} +
{A \over 12 \bkt{2 \pi t}^{\tfrac D2-1}} {1-n^2 \over n}.
\ee
 \section{Second quantization from the first quantization}\label{app:motivation}
 In this appendix we briefly motivate the form of the SFT action in the light-cone gauge. The discussion here is inspired 
by~\cite{Zwiebach:2004tj}(see sec. 11.4). 
 
 First we briefly recap some elementary aspects of string dynamics. 
 Motion of relativistic strings in $ \mathbb{R}^{1,D-1}$ is described by the maps $X^\mu\bkt{\tau,\sigma}$ from a two dimensional 
world sheet to the target space $ \mathbb{R}^{1,D-1}$ subject to the action functional
 \be
 {\rm I}=-{1\over 2\pi \alpha'} 
 \int d\tau d\sigma 
 \sqrt{\bkt{\p_\tau X^\mu \p_\sigma X_\mu}^2-{\bkt{\p_\tau X}}^2 \bkt{\p_\sigma X}^2}.
 \ee
 
 The worldsheet is parametrized by $\bkt{\tau,\sigma}$. The above action is invariant under reparameterization of the worldsheet. 
For open string we take $\sigma\in\sbkt{0,\pi}$ and for closed strings $\sigma\in\sbkt{0,2\pi}$  with identification 
$\bkt{\tau,\sigma}\sim 
\bkt{\tau,\sigma+2\pi}$. The conserved current associated with a constant shift in $X^\mu$ is $\PP_\mu^a={\p 
\LL\over \p\bkt{\p_a X^\mu}}$, $a=\tau,\sigma$ denotes wolrdsheet coordinates. The integral of the $\tau$ component over 
$\sigma$ 
give conserved charges $p^\mu$. The equation of motion is
 \be
 \p_a \PP_\mu^a=0. 
 \ee 
 In light-cone gauge, the $\tau$ parameterization is fixed by choosing
 \be
 X^+=\beta\alpha' p^+ \tau \qquad \beta=2 (1)\text{ for open(closed) strings.}
 \ee
 The $\sigma$ parameterization is fixed by demanding the constancy of $\PP^{\tau, +}$ along the string, i.e., 
 \be
 p^+={2\pi\over \beta} {\PP^{\tau,+}}.
 \ee
 Conservation of $p^+$ and equations of motion then imply that ${\PP}^{\sigma,+}$ is also constant along the string and it can be 
set equal to zero. This leads to the constraint 
 \be
 \p_\tau X^\mu \p_\sigma X_\mu=0 
 \ee
 which alongwith the $\sigma$-parameterization gives the constraint
 \be
 {\p_\tau X }^2+\bkt{\p_\sigma X}^2.
 \ee
These two constraints can be combined as
 \be \label{eq:constraint}
 \bkt{\p_\tau {X}^\mu\pm \p_\sigma X^\mu} \bkt{\p_\tau {X}_\mu\pm \p_\sigma X_\mu}=0,
 \ee
 which can be used to determine $X^-\bkt{\tau,\sigma}$ in terms of $X^{I}\bkt{\tau,\sigma}$ up to a constant. The equations of 
motion for the transverse coordinates are
 \be
 {\p_\tau^2{X}^I-\p_\sigma^2 X{}^I}=0.
 \ee
 For open strings with free end-points, these are solved by
 \be
 X^I\bkt{\tau,\sigma}= x^I+ 2\alpha' p^I \tau +\io\sqrt{2\alpha'} \sum_{n\in \mathbb{Z}\backslash \{0\}} {1\over n} \alpha_n^ I e^{-\io 
n \tau}\cos n\sigma.
 \ee
We rearrange various coefficients in the above expansion and define
\be
\begin{split}
X^I\bkt{\sigma}=X^{I}\bkt{0,\sigma}\equiv&\  x^I+\sqrt{2} \sum_{n=1}^{\infty} x_n^I \cos n\sigma, \\
P^I\bkt{\sigma}={1\over 2\pi \alpha'} \dot{X}^{I}\bkt{0,\sigma}\equiv & {1\over \pi}\bkt{p^I+\sqrt{2} \sum_{n=1}^\infty  
p_n^I \cos n\sigma
}.
\end{split}
\ee
 So classically the motion of open strings after gauge fixing is described by $\bkt{X^I\bkt{\tau,\sigma}, x^-, p^+}$ or equivalently 
$\bkt{X^I\bkt{\sigma}, 
P^I\bkt{\sigma}, x^-, p^+}$. We can quantize the system by imposing commutation relation between 
canonical pairs.
 \be
 \sbkt{X^I\bkt{\sigma}, P^J\bkt{\sigma'}}=\io \delta^{I J} \delta\bkt{\sigma-\sigma'},\qquad \sbkt{x^-,p^+}=-\io.
 \ee
  A complete set of commuting observables is $\bkt{p^+, P^I\bkt{\sigma}}$ or $\bkt{x^-, X^I\bkt{\sigma}}$.  However, none of 
these sets commute with the Hamiltonian
 \be
 H=\pi\alpha'\int _0^\pi d\sigma \bkt{\PP^I\PP^I+{1\over\bkt{2\pi \alpha'}^2} X^{I'} X^{I'}}.
 \ee
 We choose to label the states of the quantum string as
  \be
 \ket{x^-, X^I\bkt{\sigma}}.
 \ee
 Then a generic superposition
 \be
 \ket{\Phi}=\int d x^- \sbkt{ D X^I\bkt{\sigma} }\Phi\bkt{x^+,x^-,X^I\bkt{\sigma}} \ket{x^-, X^I\bkt{\sigma}}
 \ee
 satisfies the Schrodinger equation 
 \be
 \io {\p\over \p \tau} \ket{\Phi}= H\ket{\Phi},
 \ee
 if the the functional $\Phi\bkt{x^+,x^-,X^I\bkt{\sigma}}$ satisfies the equation
 \be
 \p_+\p_- + \frac{\pi}{2} \int_0^\pi d\sigma \bkt{-\frac{\delta}{\delta X^I\bkt{\sigma}}\frac{\delta}{\delta X^I\bkt{\sigma}} +\frac{1}{4 
\pi^2\alpha'^2} X'{}^I X'{}^I}
 \Phi\bkt{x^+,x^-,X^I\bkt{\sigma}}=0.
 \ee
 We can now proceed to `second quantization' in which we quantize the field $\Phi\bkt{x^+,x^-,X^I\bkt{\sigma}}$ so that it 
satisfies the above equation on-shell. This would lead to a quantum theory of string-field operators and states with multiple 
strings. It is now obvious how to write an action which gives the above equations of motion. A similar analysis can be carried out 
for closed strings with appropriate modifications. 

\small
\bibliographystyle{klebphys2} 
\bibliography{EErefs}  
 
\end{document}